\documentclass[%
 aip,
 jmp,%
 amsmath,amssymb,
 reprint,%
]{revtex4-1}

\usepackage{float}
\usepackage{multirow}
\usepackage{xcolor}
\usepackage{graphicx}
\usepackage{dcolumn}
\usepackage{bm}
\usepackage{mathtools}

\begin{document}

\preprint{AIP/123-QED}

\title[Partition-DFT on the Water Dimer]{Partition-DFT on the Water Dimer}

\author{Sara G\'omez}

 \affiliation{Grupo de Qu\'imica--F\'isica Te\'orica, Instituto de Qu\'imica,
   Universidad de Antioquia UdeA, Calle 70 No. 52-21, Medell\'in, Colombia.}

\author{Jonathan Nafziger}%
\affiliation{Department of Chemistry, Purdue University,  560 Oval Drive, West Lafayette, Indiana 47907, USA}%

\author{Albeiro Restrepo}
 \affiliation{Grupo de Qu\'imica--F\'isica Te\'orica, Instituto de Qu\'imica,
   Universidad de Antioquia UdeA, Calle 70 No. 52-21, Medell\'in, Colombia.}

\author{Adam Wasserman}
 \affiliation{Department of Chemistry, Purdue University,  560 Oval Drive, West Lafayette, Indiana 47907, USA}%
\affiliation{Department of Physics and Astronomy, Purdue University, 525 Northwestern Avenue, West Lafayette, Indiana 47907, USA}
 \email{awasser@purdue.edu}

\date{\today}

\begin{abstract}
As is well known, the ground-state symmetry group of the water dimer switches from its equilibrium $C_{s}$-character to $C_{2h}$-character as the distance betweeen the two oxygen atoms of the dimer decreases below $R_{\rm O-O}\sim 2.5$ \AA{}. For a range of $R_{\rm O-O}$ between 1 and 5 \AA{}, and for both symmetries, we apply Partition Density Functional Theory (PDFT) to find the unique monomer densities that sum to the correct dimer densities while minimizing the sum of the monomer energies.  We calculate the work inovolved in deforming the isolated monomer densities and find that it is slightly larger for the $C_s$ geometry for all $R_{\rm O-O}$. We discuss how the PDFT densities and the corresponding partition potentials support the orbital-interaction picture of hydrogen-bond formation. 

\end{abstract}

\pacs{Valid PACS appear here}
\maketitle

\section{\label{sec:intro}Introduction}

Partition density-functional theory  (PDFT) \cite{CW07, wasserman_2010, nafziger_2014} is a reformulation of DFT in which the total ground-state energy and density of a molecular system (molecules, clusters) are found indirectly, but in principle {\em exactly}, via self-consistent calculations on isolated fragments. In addition to sharing many of the appealing features of density-based embedding methods \cite{WNJK17}, PDFT can be used to formulate chemical-reactivity theory (CRT) without the inconsistencies of previous formulations \cite{CW07}, and it can be used to circumvent some of the limitations of approximate exchange-correlation functionals. For example, it was recently shown how a very simple approximation in PDFT can fix almost entirely the delocalization and static-correlation errors of approximate DFT calculations on stretched molecules \cite{NW15}.


At present, PDFT has been applied only to diatomic molecules \cite{nafziger_2011} (H$_2^+$, H$_2$, He$_2$, Be$_2$, LiH, Li$_2$), and model chains of hydrogen atoms containing non interacting electrons \cite{CWB07,wasserman_2010, TNW12, nafziger_2014}. We apply it here for the first time to a molecular cluster.  Our goals are: (1) To demonstrate convergence of the PDFT equations for a 2-fragment molecular cluster when a popular hybrid exchange-correlation functional and gaussian basis set are used; (2) to demonstrate convergence of the PDFT equations as the ground-state symmetry group of the water dimer changes with the separation between the monomers; and (3) to illustrate some of the chemical interpretations that can be drawn from a PDFT calculation.  

PDFT has much in common with related density-based embedding methods, such as Subsystem-DFT  \cite{JN14}.  However, there is an important difference.  In standard Subsystem-DFT each fragment has its own embedding potential independent of the other fragments while in PDFT all fragments share a global partition potential.  At first glance, the Subsystem-DFT picture may seem more intuitive because it is surprising that the same potential could correctly deform both fragments so that they add to yield the supermolecular density.  It turns out that it is not only possible for a single global potential to acomplish this task, but it forces the solution to be unique.  While exact Subsystem-DFT calculations which can reproduce KS-DFT calculations via fragment calculations are possible and have been done \cite{FJNV10,JBM11}, the resulting potentials and fragment densities depend either on the choice of frozen density, or on the initial guess in the case that freeze-thaw cycles are used.  In this work, the fragment densities and the partition potential are unique \cite{CW06}.  In the work of Huang and Carter \cite{HPC11}, uniqueness is imposed as an explicit constraint. For fragments with integer numbers of electrons, as here, their method is equivalent to PDFT. However, the calculations of ref.\cite{HPC11} were performed using a plane-wave basis and it is of interest to examine convergence with gaussian basis sets.

The water dimer is of course an extremely well-studied system\cite{reimers_1982, smith_1990, halkier_1997, burnham_2002, bartha_2003, dyke_exp_water_dimer_1977, curtiss_1979, odutola_dyke_1980, saykally_2001, saykally_2015,chemists_guide_dft}. It is the building unit for larger water clusters and the archetypal example of a hydrogen bond, which could be characterized as a medium-strength intermolecular interaction. Hydrogen bonds in water lead to a large number of unusual macroscopic properties\cite{w4, w5, w6, w7} (when compared to molecules of similar structure or mass\cite{h2s}) that have profound implications for the regulation of the temperature on earth,
the ability of life to thrive under frozen and extreme environments  and, more generally,
life as we know it \cite{w6}.

\begin{figure}
\begin{center}
{\hspace*{30pt} \large $\delta^-   \hspace*{50pt} \delta^+ \hspace*{20pt} \delta^- $\\}
\includegraphics[scale=0.06]{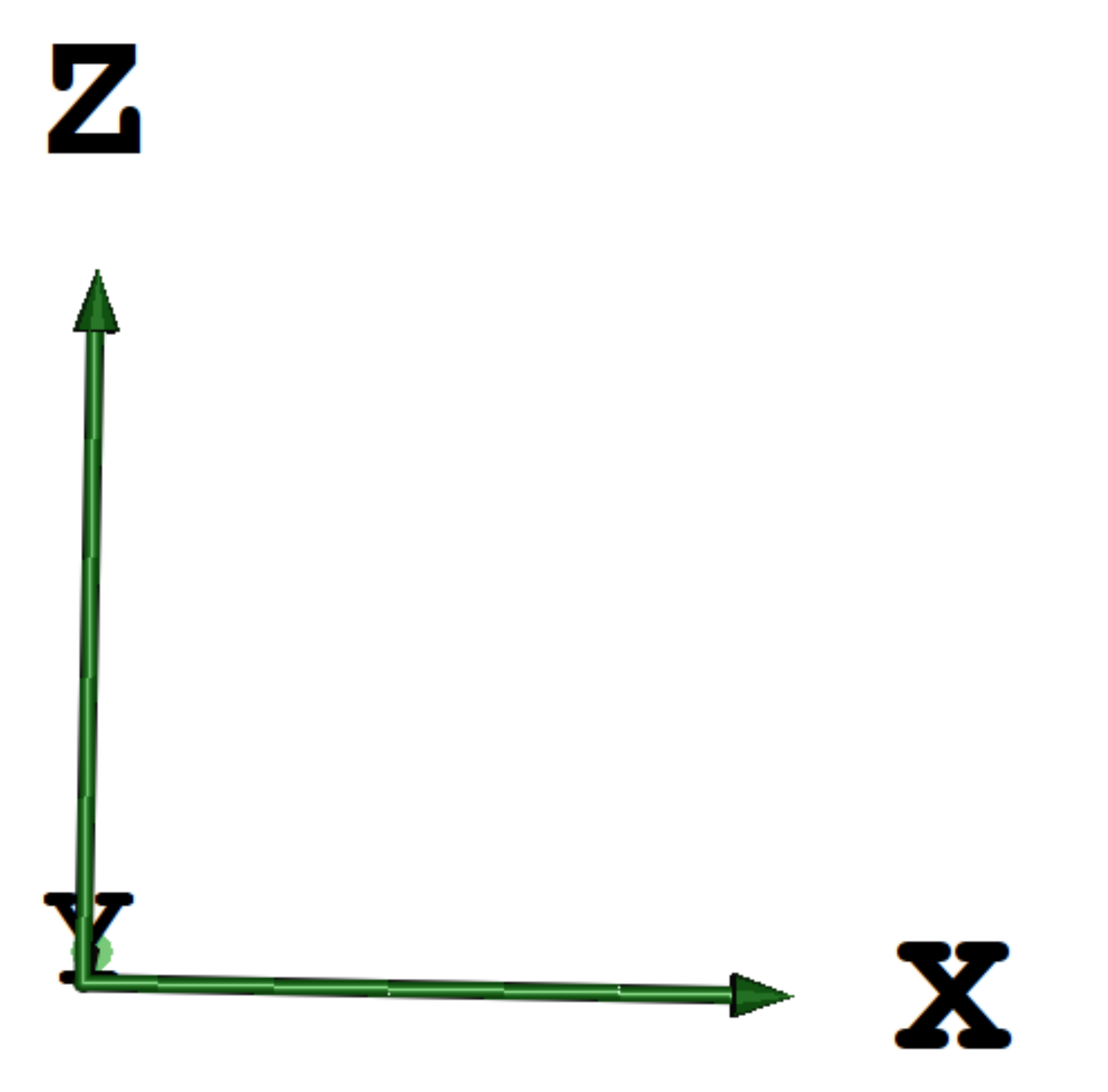}\includegraphics[scale=0.16]{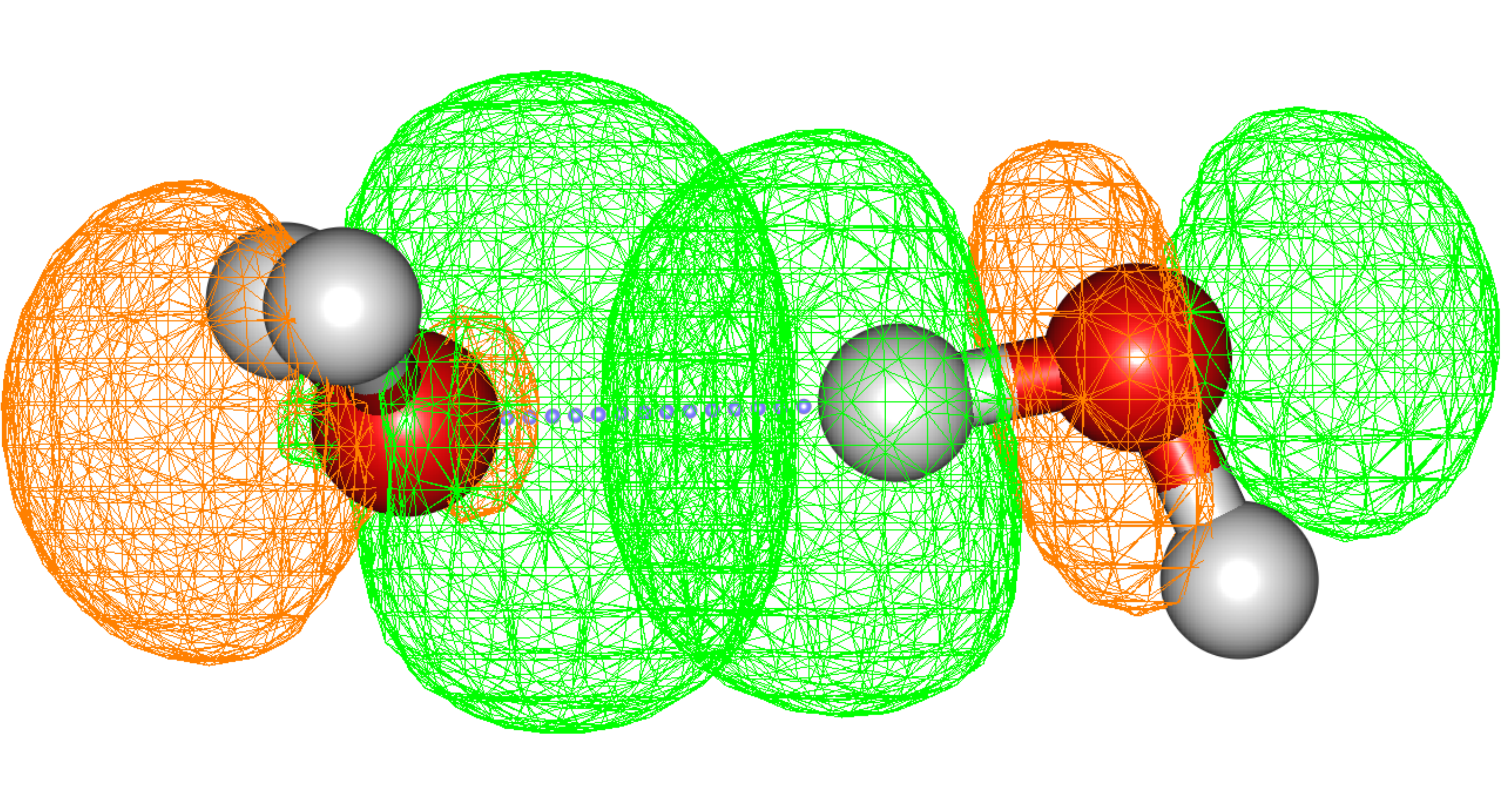}\\
\textit{$C_s$ symmetry}\\
\bigskip
\includegraphics[scale=0.06]{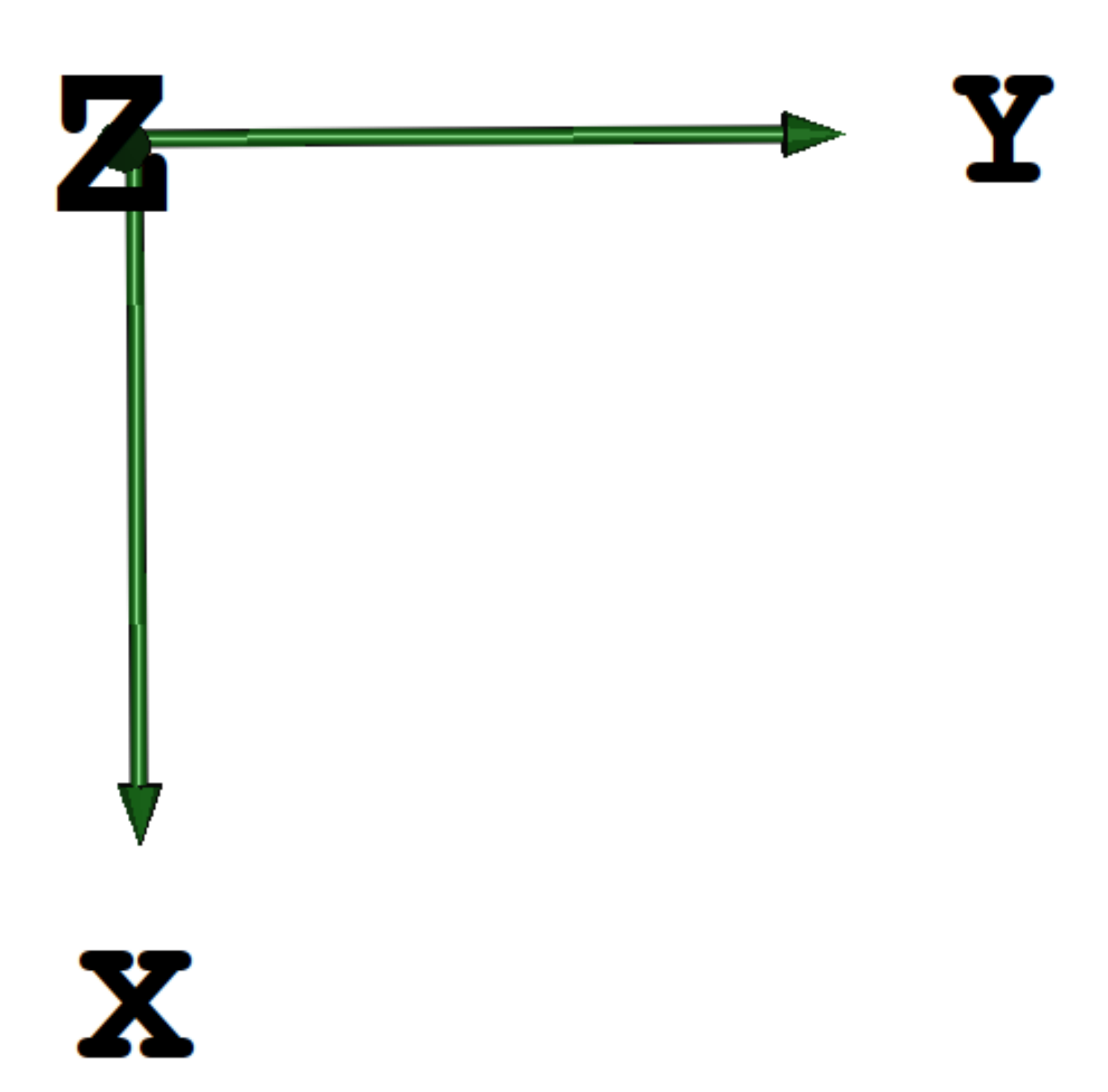}\includegraphics[scale=0.16]{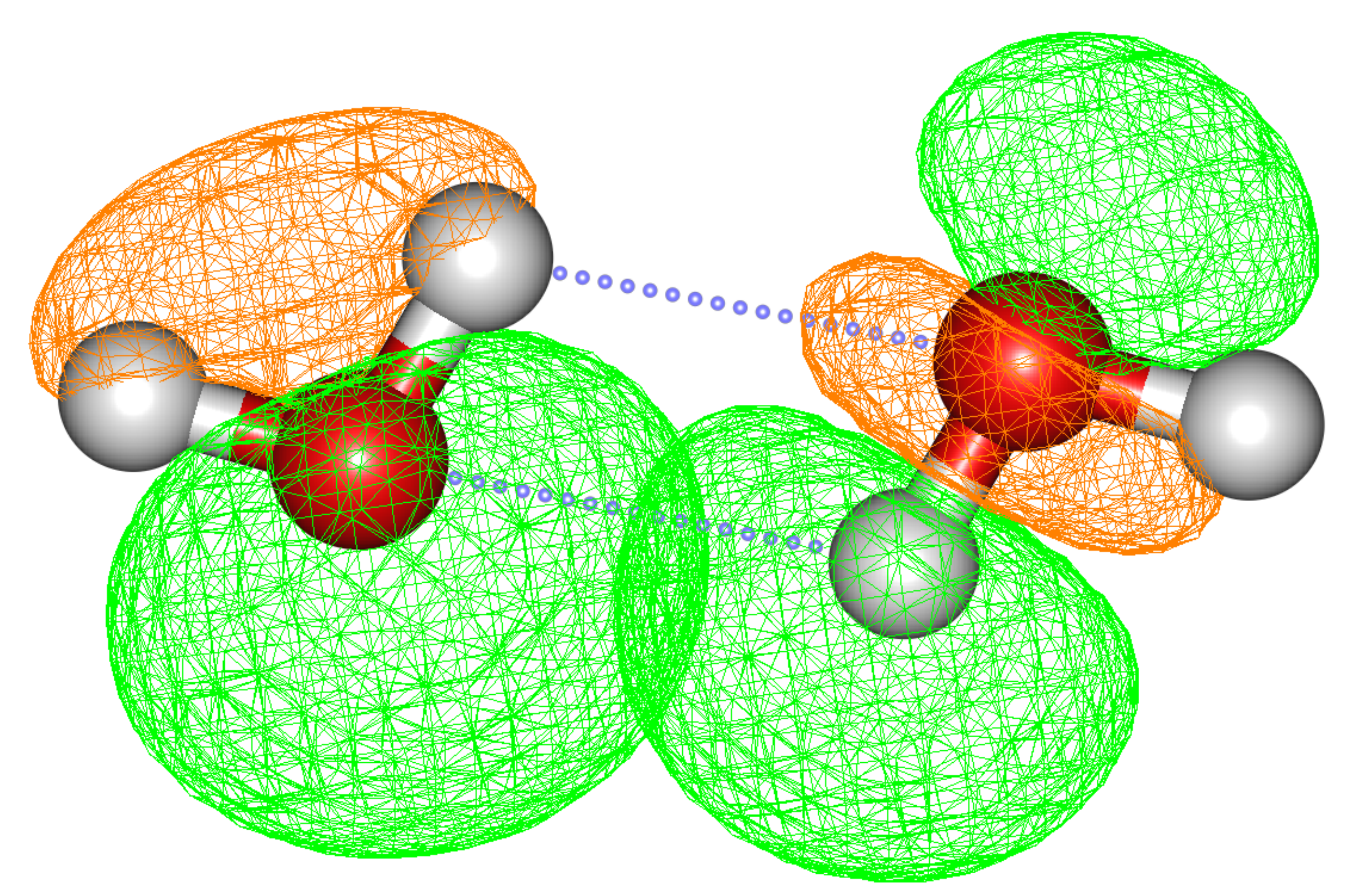}\\
\bigskip
\textit{$C_{2h}$ symmetry}
\end{center}
\caption{$C_s$ (top panel) and $C_{2h}$ (bottom panel) structures for the water dimer. In the $C_s$ cluster, the unit to the left is the {\em acceptor} of the hydrogen bond, $A$, the unit to the right is the {\em donor} of the hydrogen bond, $D$. This structure has been detected via microwave spectroscopy.\cite{dyke_exp_water_dimer_1977} Two views of the hydrogen bond are portrayed: classical electrostatic and $\left(n_{\rm O}\rightarrow\sigma^*_{\rm O-H} \right)$ orbital interaction.}
\label{f:cs_water_dimer}
\end{figure}

Currently there are two ways of understanding the stabilization in hydrogen bonds of the general Y$\cdots$H--X type, with X and Y being electronegative atoms (oxygens in the case of the water dimer). In the traditional description of electrostatic interaction between partial charges, illustrated at the top panel of Figure \ref{f:cs_water_dimer} for the lowest energy configuration of the water dimer at its equilibrium $C_s$-geometry, the X--H unit is termed the {\em donor} ($D$) and the Y unit is called the {\em acceptor} ($A$) of the hydrogen bond. In the orbital-interaction picture suggested in the classic work by Reed and Weinhold,\cite{weinhold_1988} (Figure \ref{f:cs_water_dimer}), 
charge is transferred from a non-bonding electron lone pair in the oxygen atom in the acceptor unit to the antibonding orbital in the donor molecule, $\left(n_{\rm O}\rightarrow\sigma^*_{\rm O-H} \right)$. At small separations between fragments, it is known that the ground-state geometry changes to $C_{2h}$, so $A$ and $D$ become equivalent, bonded via {\em two} hydrogen bonds. \cite{burnham_2002}

After briefly reviewing the theoretical background and computational methods in Secs. II and III,  we present in Sec. IV the PDFT energies, densities, and partition potentials, and discuss implications on our chemical understanding of the hydrogen bond.

\section{\label{sec:theory}Theoretical Background}
The formalism of partition density functional theory has been described before \cite{wasserman_2010, nafziger_2014}. Here, we merely expose the most important aspects, writing down the equations for the specific case of the water dimer.

In PDFT, $\phi_{i,\alpha}({\bf r})$, the orbitals for each $\alpha$-fragment, containing a total of $p_\alpha$ electrons, are self-consistently obtained from corresponding one-particle KS fragment equations. In our case, the most effective partition is to consider each water molecule as a fragment, the acceptor ($A$) and donor ($D$), each one containing 10 electrons. The $i$-th orbital in the acceptor fragment satisfies
\begin{equation}
 \left\{ -\frac{1}{2}\nabla^2+ v_{{\rm eff}, A}\left[n_{A}\right]\left(\mathbf{r}\right)+v_p\left(\mathbf{r}\right)  \right\}\phi_{i,A}\left(\mathbf{r}\right) = \epsilon_{i,A}\phi_{i,A}\left(\mathbf{r}\right).
 \label{eq:ksf}
\end{equation}
Replacing $A$ by $D$ yields an identical set of equations for the donor fragment. Here, $v_{{\rm eff},A}$, the effective potential for the hydrogen bond-acceptor unit, is the sum of the Hartree $v_{\rm H}(\mathbf{r})$, exchange--correlation $v_{\rm XC}(\mathbf{r})$, and external (nuclear) $v_A(\mathbf{r})$ potentials,
\begin{equation}
 v_{{\rm eff},A}\left[n_A\right]\left(\mathbf{r}\right)=v_{\rm H}\left[n_A\right]\left(\mathbf{r}\right) + v_{\rm XC}\left[n_A\right]\left(\mathbf{r}\right) + v_A\left(\mathbf{r}\right)
 \label{eq:ep}
\end{equation}
The set of occupied KS orbitals obtained from equation (\ref{eq:ksf}) determine the fragment densities according to
\begin{equation}
 n_{A}\left(\mathbf{r}\right) = \sum_{i}^{occ}\left|\phi_{i,A}\left(\mathbf{r}\right)\right|^2
 \label{eq:fd}
\end{equation}
The total energy of the water dimer is obtained as
\begin{equation}
\begin{multlined}
 E[n_A,n_D] = 
  E_A[n_A]+E_D[n_D] +E_p[n_A,n_D]
  \end{multlined}
 \label{eq:te}
\end{equation}
where $E_p[n_A,n_D]$ is the partition energy, and $E_A[n_A]=T_s[n_A]+E_{\rm H}[n_A]+E_{\rm XC}[n_A]+\int d{\bf r} v_A({\bf r}) n_A({\bf r})$ does not include the energy contribution from the partition potential. 

We are also interested in how the fragment densities deform from the isolated monomer densities.  We denote $E_{A}^0=E_{D}^0$ the energy of the isolated water monomer, which allows us to define the {\em preparation energy} of the acceptor unit as the energy needed to take it from the geometry and charge distribution of the isolated water monomer to the acceptor geometry and charge distribution in the dimer,
\begin{equation}
 E_{{\rm prep},A}[n_A] = E_A[n_A]-E_A^0
 \label{eq:eprep}
\end{equation}
with a similar definition for the donor. Thus, the total preparation energy for the water dimer is their sum 
\begin{equation}
 E_{\rm prep}[n_A,n_B]=E_{{\rm prep},A}[n_A]+E_{{\rm prep},D}[n_B]
 \label{eq:tpe}
\end{equation}
As is usual, the binding energy of the dimer is calculated as the difference between the energy of the dimer and the energy of the isolated monomers:
\begin{equation}
 E_{\rm bind} = E - \left(E_A^0+E_D^0\right) = E_{\rm prep} + E_{p}
 \label{eq:be}
\end{equation}

The partition potential, $v_p\left(\mathbf{r}\right)$, is common to both the acceptor and donor units and emerges as the Lagrange multiplier ensuring that the sum of the fragment densities matches the total density of the interacting dimer.
The partition potential can be related to functional derivatives of $E_p$; however, because its exact form is not known, iterative methods for optimizing $v_p$ have been developed.\cite{nafziger_2011, wasserman_2010, nafziger_2014} These involve writing $v_p$ as linear combinations of basis functions and directly optimizing the coefficients so that the the sum of fragment densities matches a precalcuated supermolecular density, $n_m$ when projected onto the basis functions of $v_p$. 

\section{\label{sec:details}Details of the calculations}
The optimization of the coefficients in the expansion for the partition potential can proceed via two algorithms.  In the simpler of the two algorithms, each coefficient representing the partition potential is updated using the equation

\begin{equation}
\delta v_{p,i} = \gamma \left(\sum_\alpha n_{\alpha,i} - n_{m,i}\right)
\end{equation}

where the $i$ index runs over the partition potential basis functions, $\alpha = A,D$, and $n_{\alpha,i}$ and $n_{m,i}$ are the basis function coefficients obtained by projecting the fragment density, $n_\alpha$ and the target supermolecular density, $n_m$, onto the partition potential basis set. $\gamma$ is a small positive constant which controls convergence.  Typically, $\gamma$ may be chosen to be around $0.05$, but depending on the system or geometry certain values of $\gamma$ will not converge.
 However, we found faster and more robust convergence by using the sum of fragment density responses, $\chi_f$, projected onto the partition potential basis.  By inverting this matrix we obtain a first order estimate of how to change the partition potential coefficients in order to zero the difference between the sum of fragment densities and the target density.
 
\begin{equation}
\delta v_{p,i} = \sum_j \chi^{-1}_{f,i,j} \left(\sum_\alpha n_{\alpha,i} - n_{m,i}\right)
\label{response_update}
\end{equation}

\begin{figure}
\scalebox{0.3}{\includegraphics{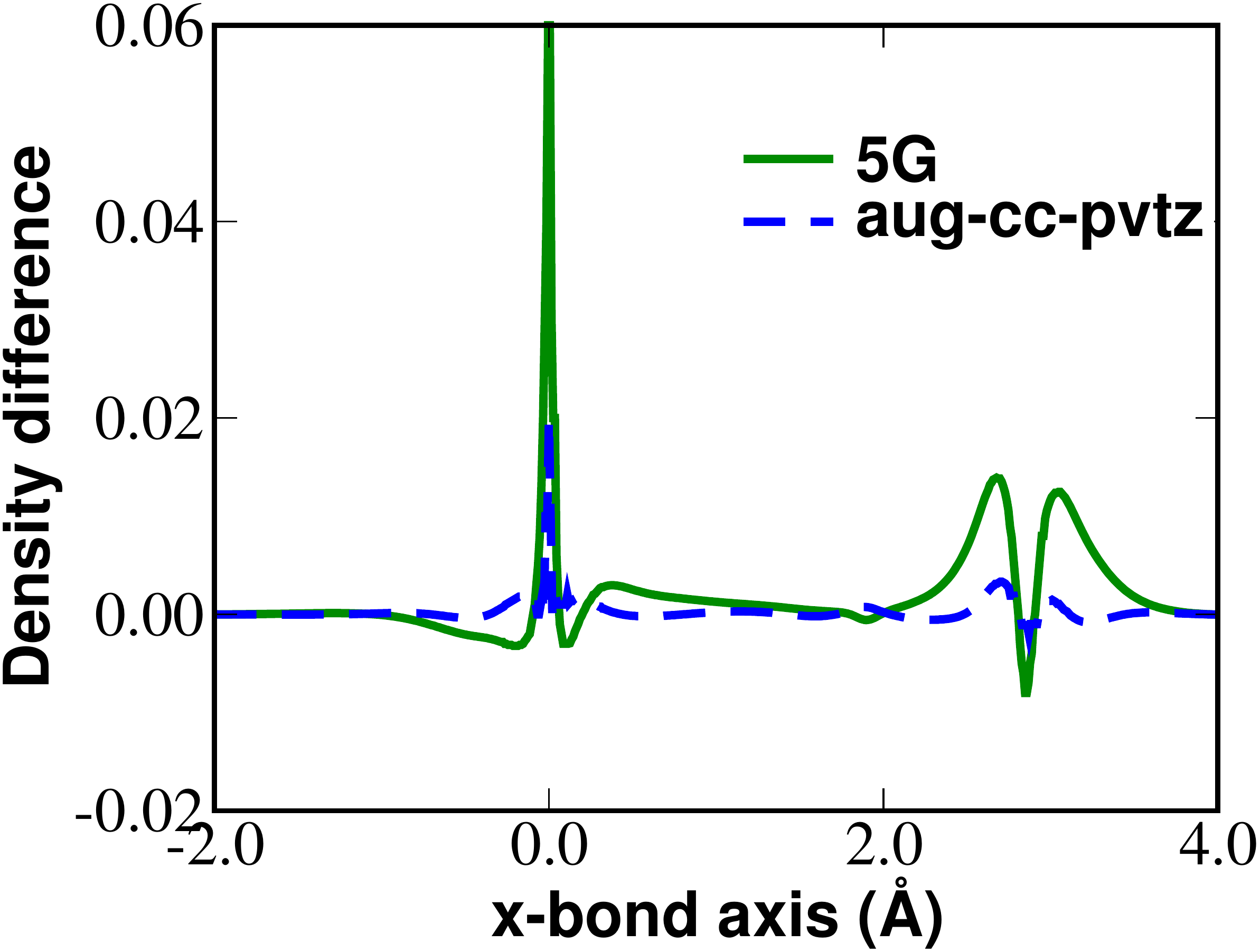}}

\caption{Difference between sum of fragment densities and target density for the equilibrium $C_s$ configuration after expanding  $v_p\left(\mathbf{r}\right)$ with 5 Cartesian Gaussians centered at each atom, and with aug--cc--pVTZ.}
\label{f:vpbasiscompare}
\end{figure}

The partition potential was expanded in terms of two different types of basis sets: 5 Cartesian Gaussian functions centered at each atom, and Dunning's aug--cc--pVTZ basis set. It can be seen from Figure \ref{f:vpbasiscompare} that expanding the partition potential over the aug--cc--pVTZ basis set allows the fragment densities to better match the total density. Therefore, this basis set is used for the rest of the calculations in this work.  All calculations were carried out using our PDFT implementation in the NWChem package.\cite{nwchem}

We separately optimized the water dimer in $C_s$ and $C_{2h}$ symmetries (Figure \ref{f:cs_water_dimer}) using second-order perturbation theory (MP2) and the B3LYP functional in conjunction with the aug-cc-pVTZ basis set. As expected, negligible differences in the geometries were obtained with the two methods. Thus, in what follows, we use the MP2 geometries. A number of rigid and relaxed scans of the separation distance between the monomers were carried out using B3LYP. In order to construct the effective potentials of equation \ref{eq:ep}, the B3LYP exchange correlation functional was used.  A few comparisons were performed using Hartree--Fock exchange as well as the LDA exchange and correlation functional.

\section{\label{sec:results}Results and Discussion}


\subsection{Energetic Analysis}
Table \ref{t:energies_cs_minimum} shows the partition and preparation energies for both monomers and both equilibrium geometries of the water dimer (Figure \ref{f:cs_water_dimer}). $E_{{\rm prep},A}$ is slightly smaller than $E_{{\rm prep},D}$. This means that the proton donor in the hydrogen bond reorganizes its charge density to a larger extent, due to the intermolecular interaction. From the orbital perspective, this makes perfect sense as this unit has to accommodate the electron charge being donated to the $\sigma_{\rm O-H}^*$ orbital.   

\begin{table}
\caption{Energies (a.u.) for PDFT (B3LYP/aug--cc--pVTZ) calculations on the $C_s$ and $C_{2h}$ equilibrium geometries of the water dimer (Figure \ref{f:cs_water_dimer}). $v_p\left(\mathbf{r}\right)$ was expanded using  aug--cc--pVTZ. $A$ and $D$ are the {\em acceptor} and the {\em donor} of the hydrogen bond (either monomer in the case of $C_{2h}$ symmetry). The energy for the isolated water monomer is $E_A^0=E_D^0=$-76.4673031738 a.u. $R_{\rm O-O} = 2.86$ \AA{} for $C_s$ and $R_{\rm O-O} = 2.76$ \AA{} for $C_{2h}$ at the equilibrium geometries. } 
\label{t:energies_cs_minimum} 
{\normalsize
\begin{center}
\begin{tabular}{c|cc|cc|} 
\cline{2-5}
\multirow{2}{*}{} & \multicolumn{2}{c|}{$C_s$}  & \multicolumn{2}{c|}{$C_{2h}$}\\ 
\cline{2-5}
& $A$ & $D$ & $A$ & $D$\\
\hline
\multicolumn{1}{|c|}{$E_{\rm prep,\alpha}\times 10^3$}&  1.12 & 1.16 & 0.33 & 0.33\\
\hline
\multicolumn{1}{|c|}{$E_{\rm prep}\times 10^3$}& \multicolumn{2}{c|}{2.28} & \multicolumn{2}{c|}{0.67}\\
\multicolumn{1}{|c|}{$E_{p}\times 10^3$}& \multicolumn{2}{c|}{-9.50} & \multicolumn{2}{c|}{-5.58}\\
\multicolumn{1}{|c|}{$E_{\rm bind}\times 10^3$}& \multicolumn{2}{c|}{-7.23} & \multicolumn{2}{c|}{-4.92}\\
\hline
\end{tabular}
\end{center}
}
\end{table}

The same conclusions are drawn from calculations employing LDA or Hartree Fock exchange. However, LDA overestimates $E_{\rm bind}$ by 2.4 kcal/mol, while HF underestimates it by 3.1 kcal/mol with respect to the experimental value (5.4 $\pm$ 0.7 kcal/mol\cite{curtiss_1979}). Thus, in what follows we use B3LYP for all calculations. 

The total preparation energy is small ($< 0.003$ a.u.). For diatomic molecules, the character of the bond seems to be related to the magnitude of the preparation energy: Nafziger and coworkers \cite{nafziger_2011} reported preparation energies for He$_2$ (van der Waals bond), LiH ionic fragments (ionic bond), LiH neutral fragments (ionic bond), and H$_2$ (covalent bond) of $5.66\times10^{-4}$, $3.44\times10^{-2}$, $5.33\times10^{-2}$, $3.26\times10^{-2}$ a.u. respectively at the equilibrium distances.  
Our results for the water dimer at the same level of theory ($2.28\times10^{-3}$ a.u.) nicely falls in the intermediate region between a long range van der Waals bond and an ionic bond. This intermediate nature of the hydrogen bond in water clusters is well documented. \cite{w5,w7,natalia} 

It can be argued that in general, in interacting systems, when the fragments retain their identities to high degrees (weak interactions), preparation energies should be smaller than in systems where the units are significantly changed from their isolated forms (medium to strong interactions). Since the preparation energy is able to describe the relative strengths of bonding, a PDFT calculation gives insight into the nature of the interaction.  

\begin{figure}\centering
\includegraphics[scale=0.35]{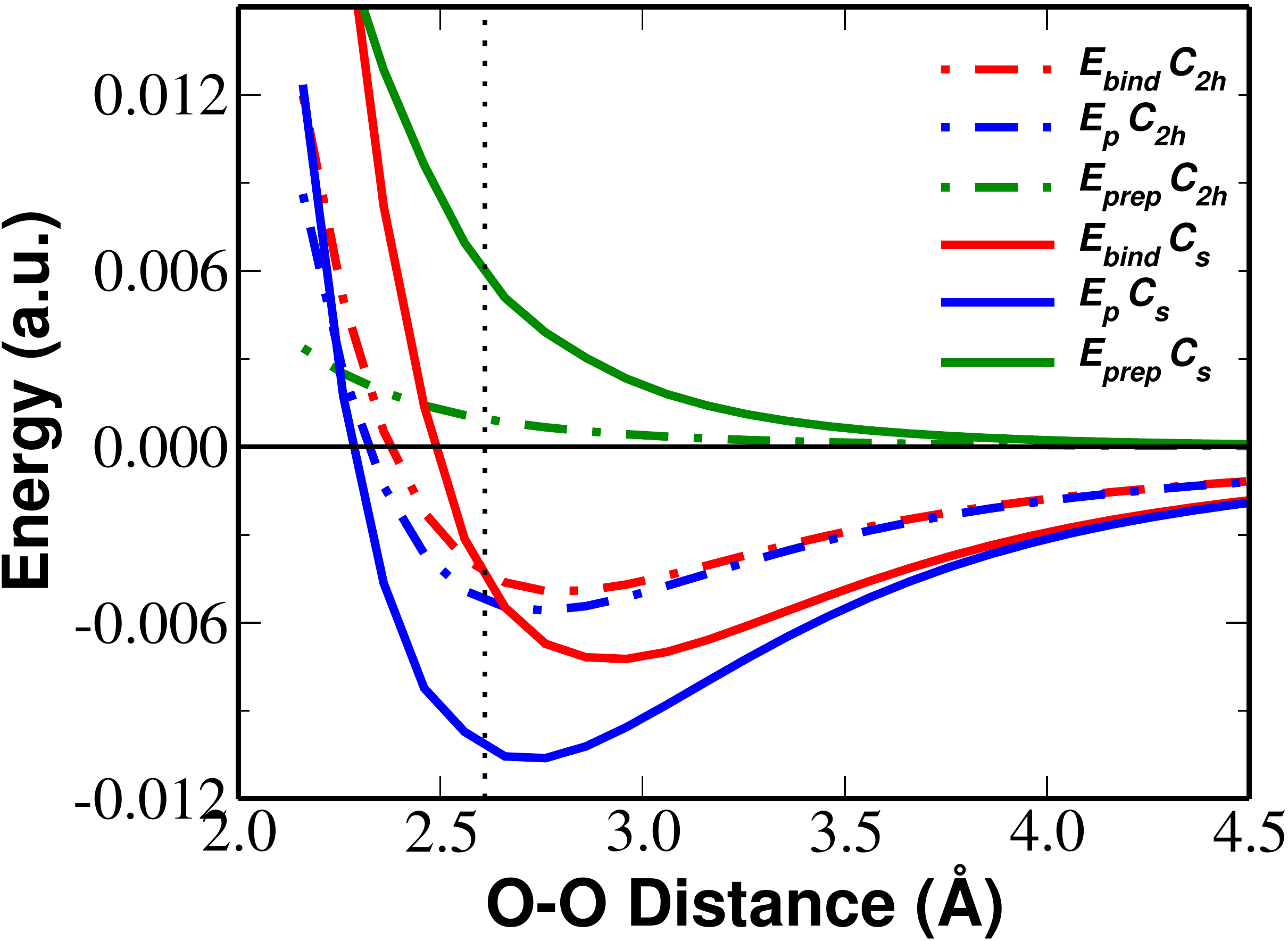}
\caption{PDFT (B3LYP/aug--cc--pVTZ) energies for the water dimer at different intermolecular distances for both $C_s$ (solid) and $C_{2h}$ (dashed) geometries. Binding energies (red) cross at $R_{\rm O-O}\sim 2.5$ \AA, indicated by the dotted vertical line. The preparation energies (green) decay fast with $R_{\rm O-O}$, and $E_{\rm prep}(C_{2h})<E_{\rm prep}(C_s)$ always. The partition energies (blue) include the non-additive nuclear-nuclear repulsion.}
\label{f:scans}
\end{figure}

Next, we focus on the changes in preparation and partition energies as a function of the intermolecular distance.  Relaxation along the $C_s$ path leads the system to change its symmetry to $C_{2h}$ at short distances as shown in Figure \ref{f:scans}. The $C_{2h}$ dimer contains doubly hydrogen bonded water molecules (both molecules act as simultaneous acceptor/donor of hydrogen bonds), while $C_{s}$ has a single hydrogen bond. This is a well known fact, pointed out among others by Burnham and Xantheas\cite{burnham_2002} who reported that the $C_{s}\rightarrow C_{2h}$ transition occurs around $R_{\rm O-O} \approx 2.50$ \AA.  While $E_{\rm bind}$ remains continous through this transition, $E_{p}$ and $E_{\rm prep}$ have a discontinuity.  However, symmetry--constrained scans of the separation distance avoid the symmetry crossing and allow us to analyze the changes in preparation and partition energies within fixed symmetries (Figure \ref{f:scans}).  Intra-fragment nuclear re-arrangement would introduce a small positive change in the preparation energy, because the fragment energy would be slightly further from the ground state of the isolated system.  It would also have a slightly larger negative effect on the partition energy for a small net lowering of the total energy.  However, these small effects are ignored due to the fixed intrafragment geometry.

The preparation energy is always positive and decays rapidly with separation.  It is interesting to note that the preparation energy of the monomers in the $C_{2h}$ isomer is significantly smaller than in the $C_s$ case ($E_{\rm prep}(C_{2h}) < E_{\rm prep}(C_{s})$). This indicates that the work involved in deforming the monomers in the $C_{2h}$ isomer with a double hydrogen bond is smaller than the work needed to deform the isolated monomers in the $C_s$ isomer with a single hydrogen bond.  This lines up with the orbital picture because as can be seen in Figure \ref{f:cs_water_dimer}, all orbitals involved have the correct symmetries and energies, but for the $C_{2h}$ case, the $n_{\rm O}\leftrightarrow\sigma^*_{\rm O-H}$ overlap is significantly smaller. In addition, for the $C_{2h}$ case, the orbitals involved in one hydrogen bond may interfere in a non--constructive way with the orbitals in the second hydrogen bond, resulting in a smaller change in fragment density. The smaller preparation energy (smaller change in fragment density) for the $C_{2h}$ isomer is also reflected in the smaller changes in the magnitudes of the dipole moments of the monomers (in fact, the changes in the $A$, $D$ monomers cancel out such that the total dipole moment remains unchanged) as compared to the monomers in the $C_s$ dimer, depicted in Figure \ref{f:dm_cs_c2h}.

\subsection{The Partition Potential}
\begin{figure}\centering
\includegraphics[width=.4\textwidth]{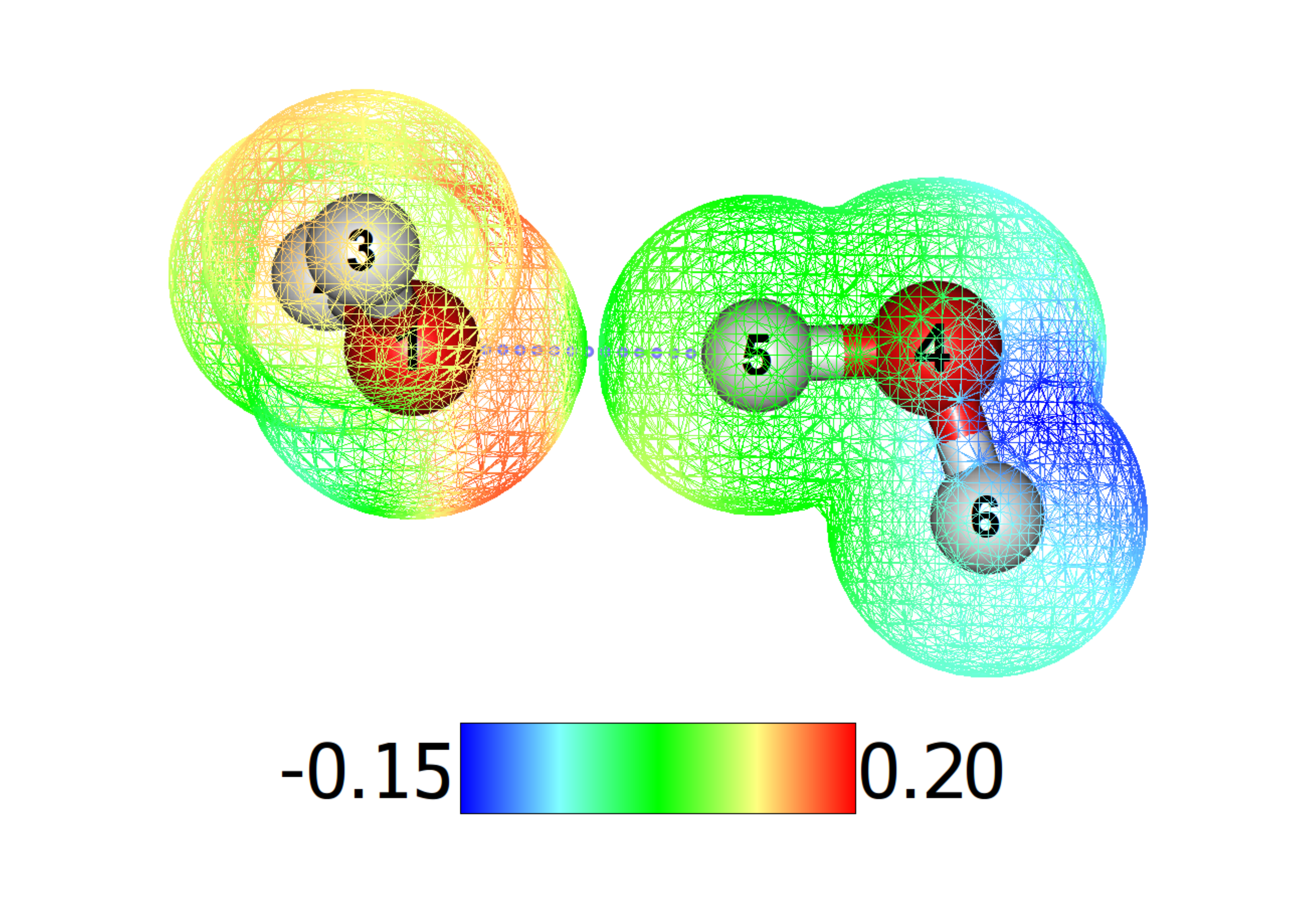}
\includegraphics[width=.4\textwidth]{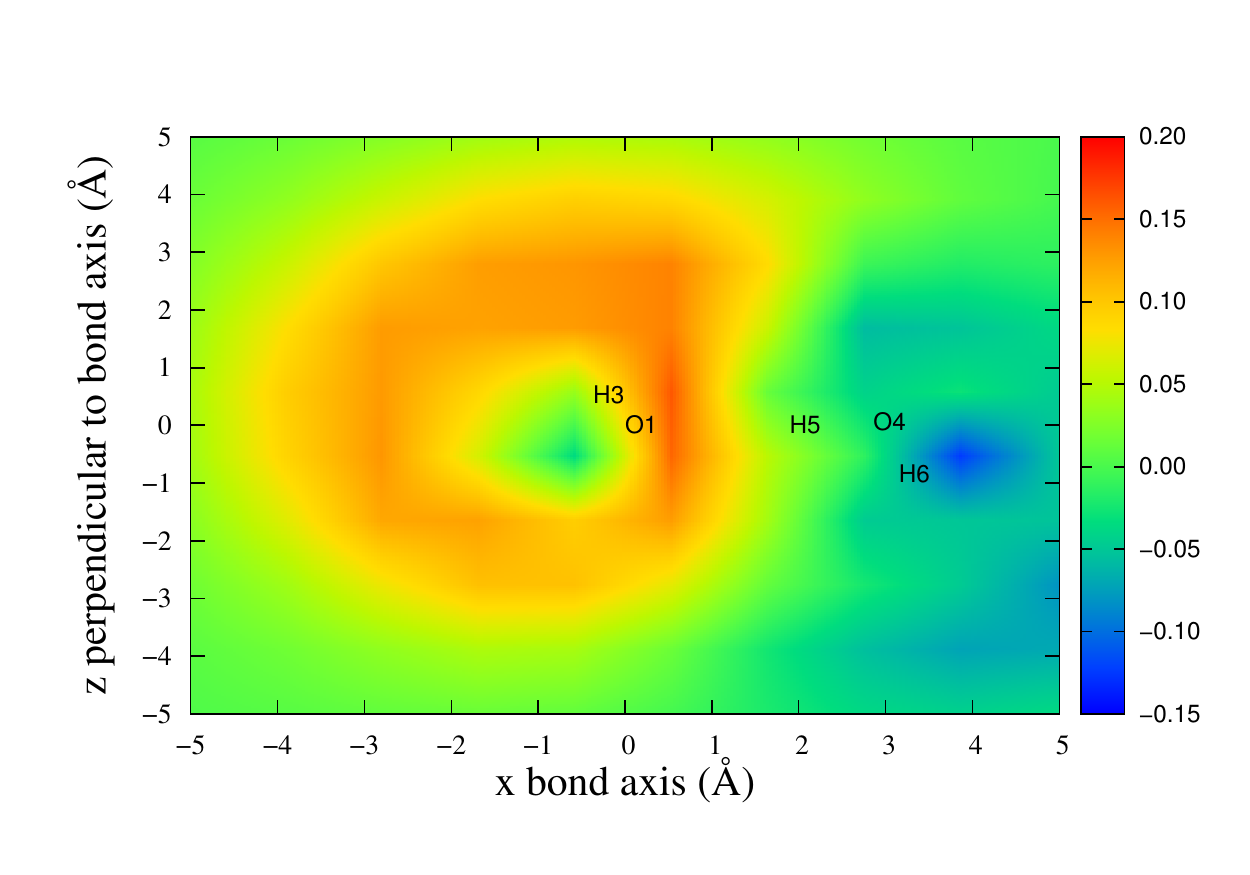}
\includegraphics[width=.4\textwidth]{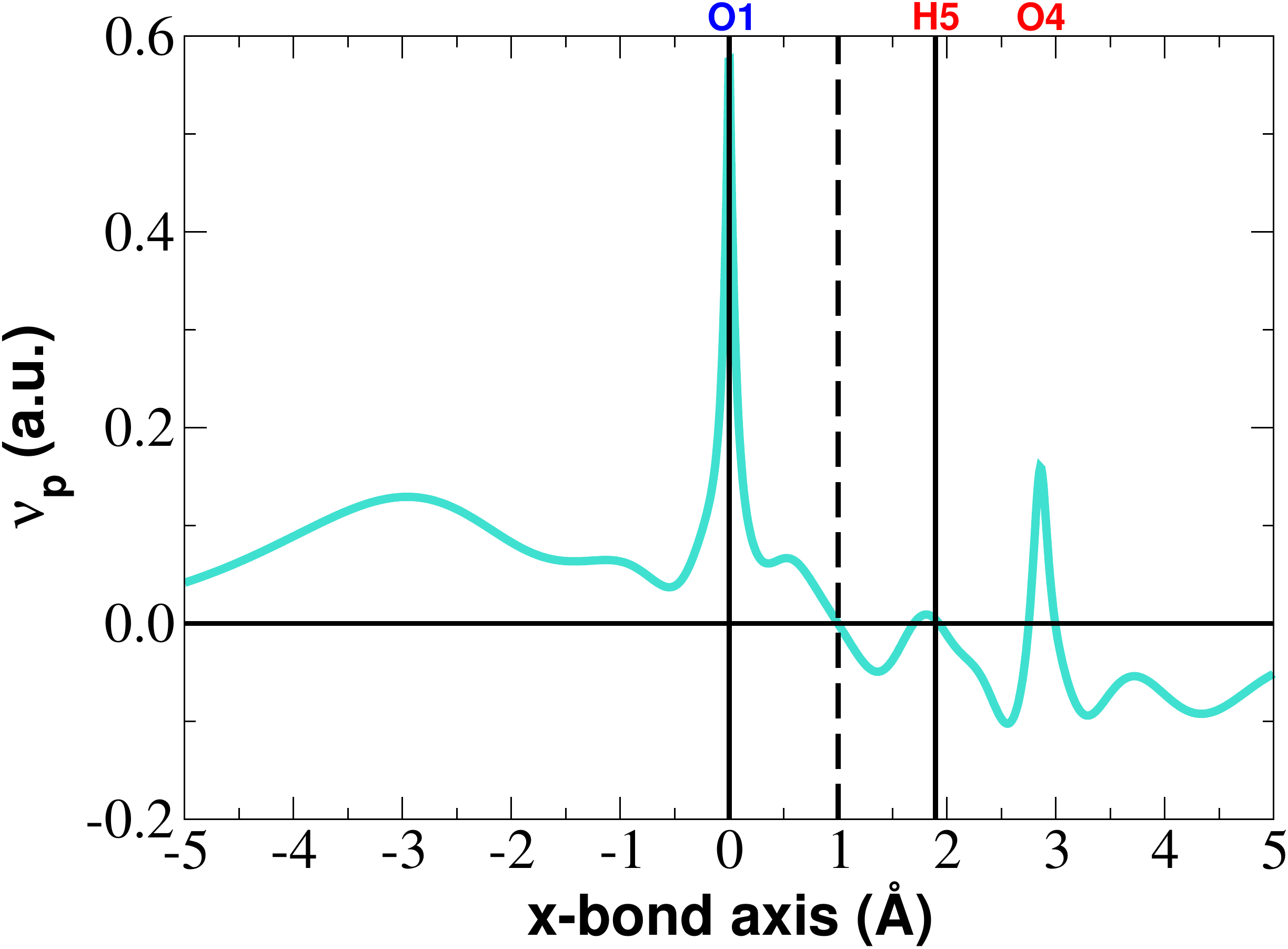}
\caption{PDFT (B3LYP) partition potential for the $C_s$ dimer at equilibrium expanded using Dunning's aug--cc--pVTZ basis set. Vertical lines in the 1D plots enclose the intermolecular bonding region.} 
\label{f:vp_plots}
\end{figure}

Figure \ref{f:vp_plots} visualizes $v_p\left(\mathbf{r}\right)$ for the $C_s$-equilibrium geometry in three different ways: a 3D surface, a 2D contour plot on the plane defined by the donor molecule that also contains the oxygen atom in the acceptor unit, and a one-dimensional cut along the bonding $x$--axis. 
Positive (red) regions of the potential are associated with charge deficiency while negative (blue) regions of the potential are associated with excess charge. If the monomers are separated by very long distances, the entire surface is green ($v_p\left(\mathbf{r}\right)=0$). What is observed in the 3D surfaces is that, in agreement with the orbital view of hydrogen bonding (Figure \ref{f:cs_water_dimer}), charge is transferred from $A$ to $D$. The 2D contour plot (aug--cc--pVTZ expansion of $v_p\left(\mathbf{r}\right)$) implies that the largest deficiency in charge appears to be at the oxygen atom in the acceptor molecule, precisely at the position of the lone pair. It is also shown in Figure \ref{f:dens_plots} that the most gain in charge in the donor molecule occurs in the region of the $\sigma_{\rm O-H}^*$ orbital. The bonding region is conveniently described by the one-dimensional plot along the bonding $x$--axis: as in the cases of the diatomic molecules mentioned above, \cite{nafziger_2011, nafziger_2014} the partition potential is somewhat diminished in the bonding region, decreasing in the O1$\rightarrow$H5 direction, which we associate with the electron flux due to the $n_{\rm O}\rightarrow\sigma_{\rm O-H}^*$ charge transfer. 

\subsection{Fragment Densities}
\begin{figure}\centering
\includegraphics[width=.30\textwidth]{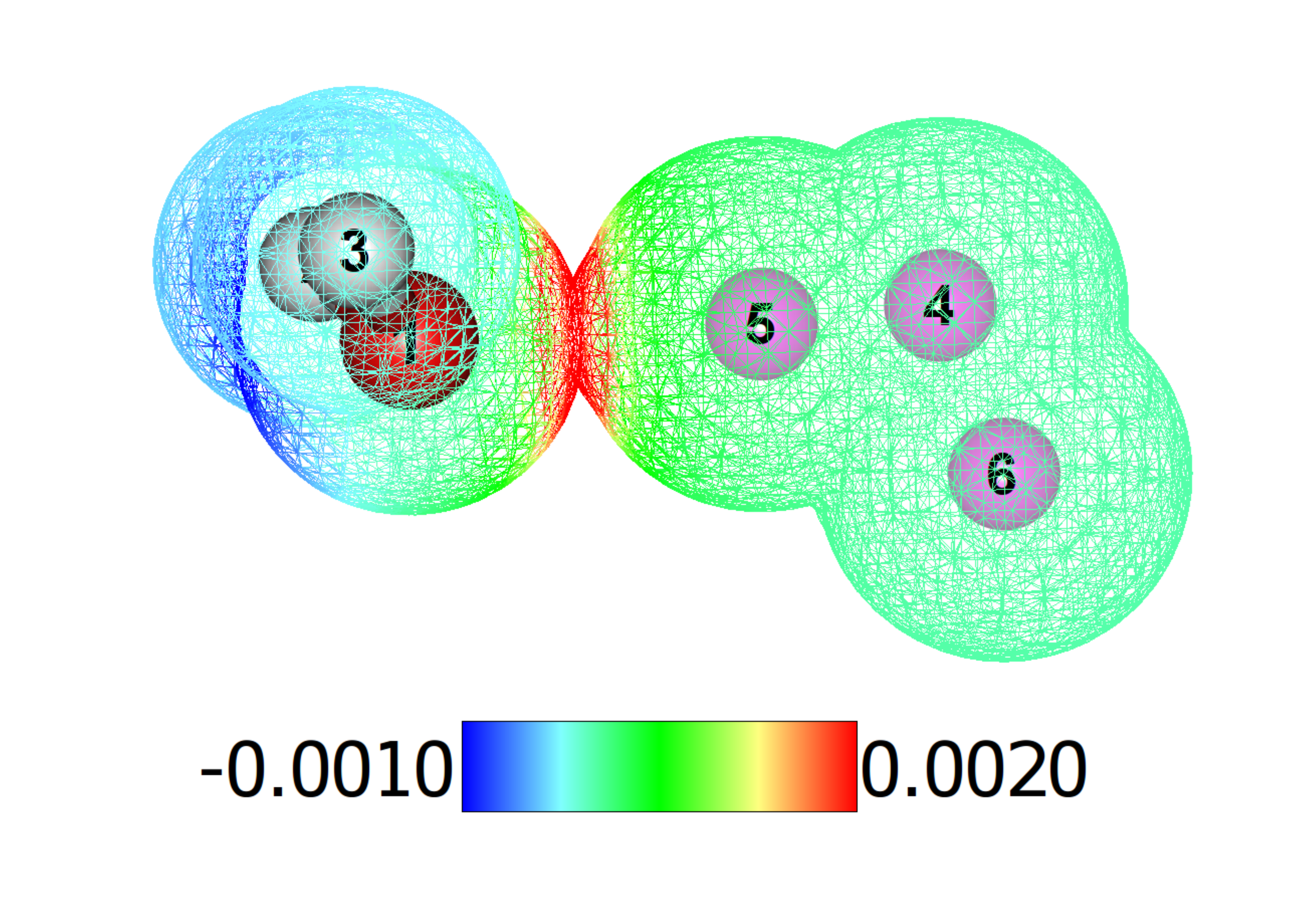}
\includegraphics[width=.30\textwidth]{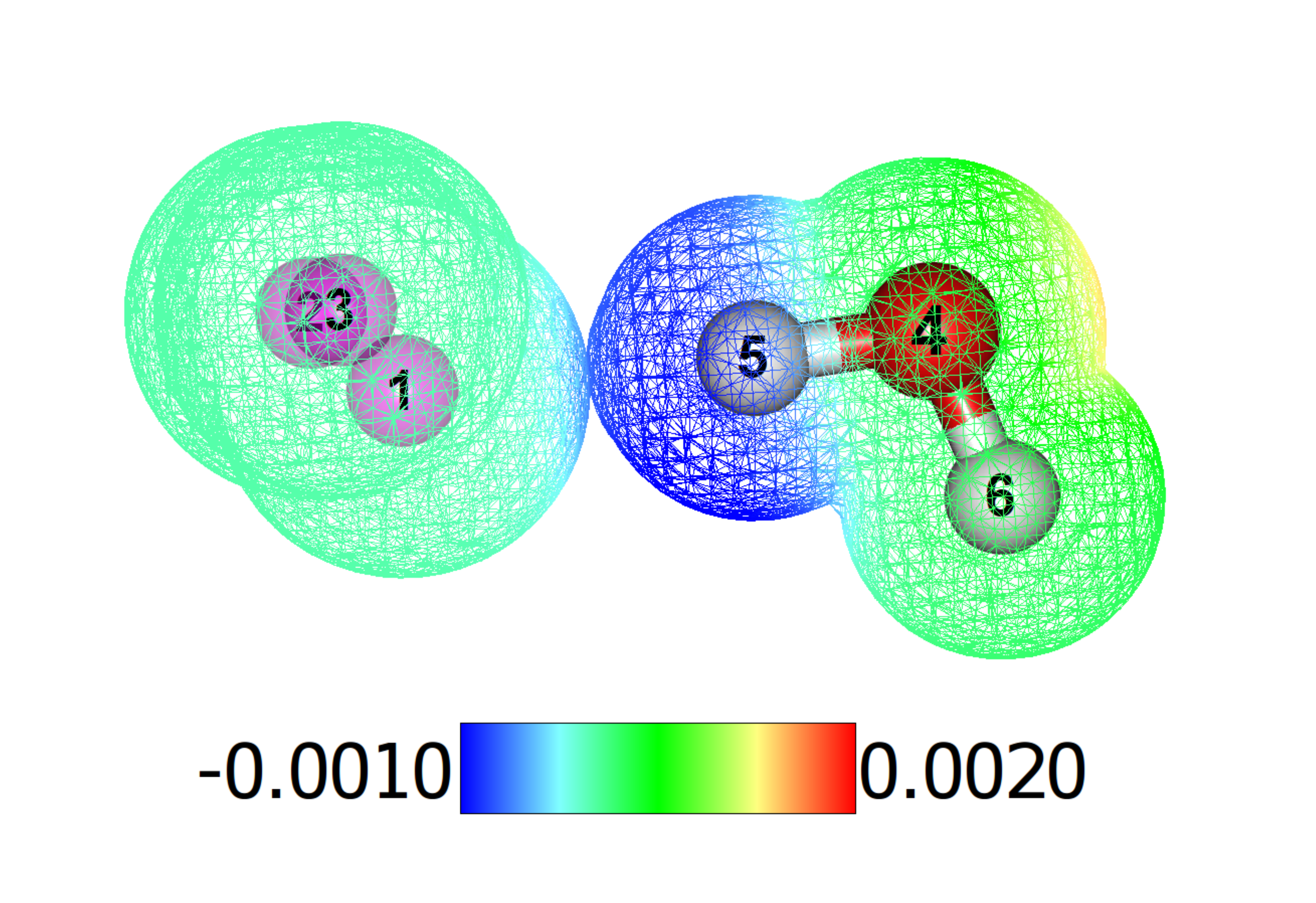}
\includegraphics[width=.4\textwidth]{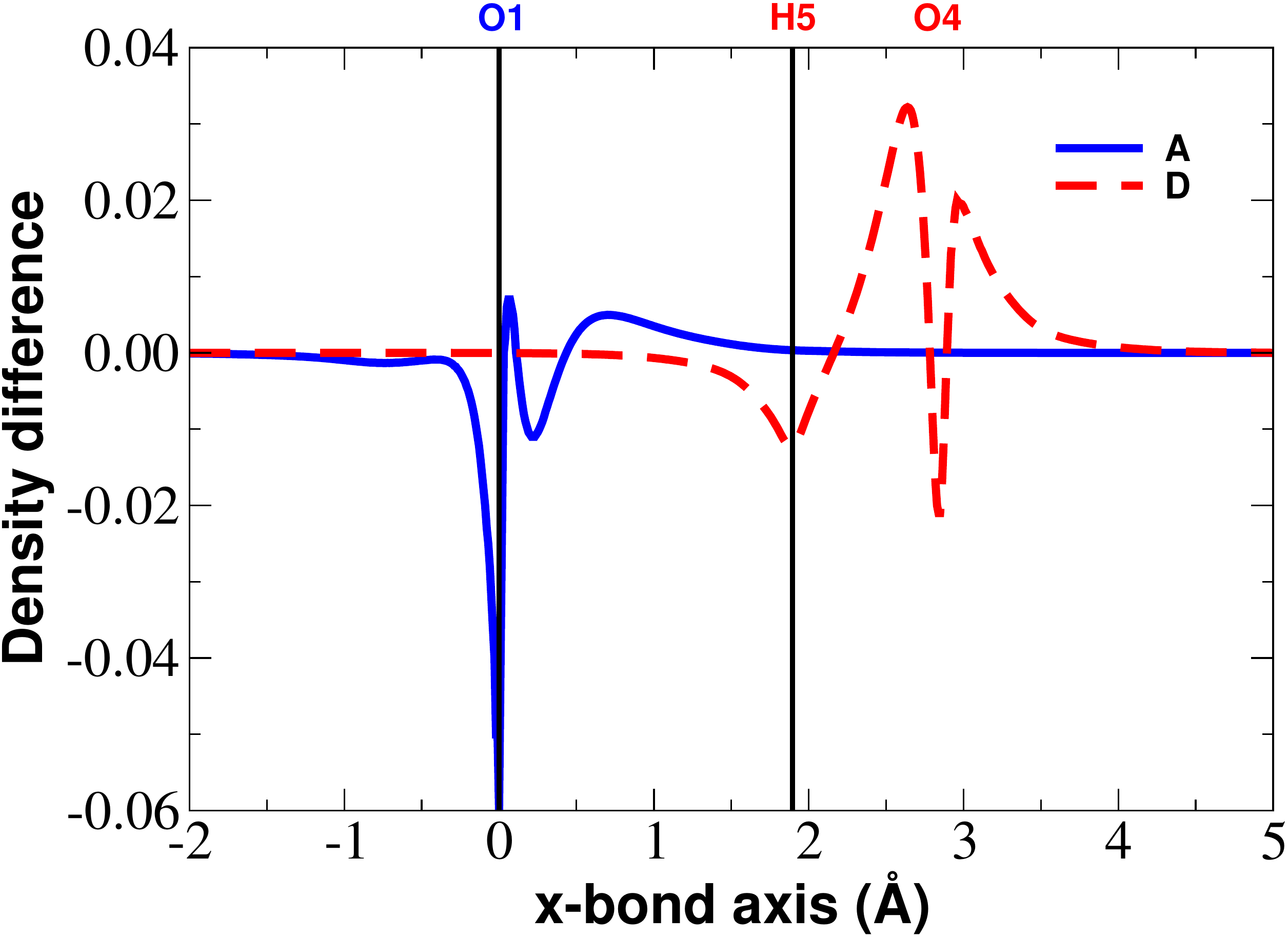}
\caption{Density differences in the monomers $A$ (top) and $D$ (middle) as compared to the original monomers for the $C_s$ dimer at equilibrium geometry. $v_p\left(\mathbf{r}\right)$ is expanded using the aug--cc--pvtz basis set. One dimensional cuts for both differences along the bonding $x$-axis is also shown (bottom). Vertical lines in the 1D plots enclose the intermolecular bonding region.} 
\label{f:dens_plots}
\end{figure}

The flux of electrons involved in the formation of the hydrogen bond is beautifully visualized using the one dimensional density difference plots along the bonding axis in Figure \ref{f:dens_plots}. It is clearly seen that the acceptor monomer loses electron density at the O1 atom, while there are two places of net electron density gain: the region associated with the intermolecular bond and most noticeably, the region corresponding to the antibonding orbital in the donor monomer, thus clearly supporting the $n_{\rm O}\rightarrow\sigma_{\rm O-H}^*$ charge-transfer picture. 
Bartha and coworkers \cite{bartha_2003} found that the total density difference changes sign several times along the hydrogen bond and related these sign changes to the electron flux associated to the intermolecular bond. 


\subsection{Dipole Moments}

 The static dipole moment of a molecule reflects a non-uniform charge distribution. Since charge distributions rearrange in the presence of electric fields, dipole moments are very sensitive to chemical environments. The changes in the dipole moments of individual monomers in a cluster encode useful information about the intermolecular interaction.  PDFT provides one way of tracking the changes of the dipole moment of a fragment within a cluster. 

 For the water monomer, the experimentally measured electric dipole is 1.855 Debyes. \cite{lovas} 
The NIST database lists 1.846 as the dipole moment calculated at the B3LYP/aug--cc--pVTZ level of theory.\cite{CCCBDB} Our calculations in this work match the experimental value $\mu_A^0 = \mu_D^0 = 1.855$ Debyes. For the water dimer, the experimental dipole moment is 2.643 Debyes. \cite{dyke_exp_water_dimer_1977} For the $C_s$ isomer,  the calculated dipole moments are also in excellent agreement with the experiment, for example, 2.632 and 2.683 Debyes have been reported at the B3LYP/aug--cc--pVTZ\cite{CCCBDB} and MP2/aug--cc--pVTZ\cite{gregory_science} levels respectively.
 
\begin{center}
 \begin{figure}\centering
\includegraphics[scale=0.30]{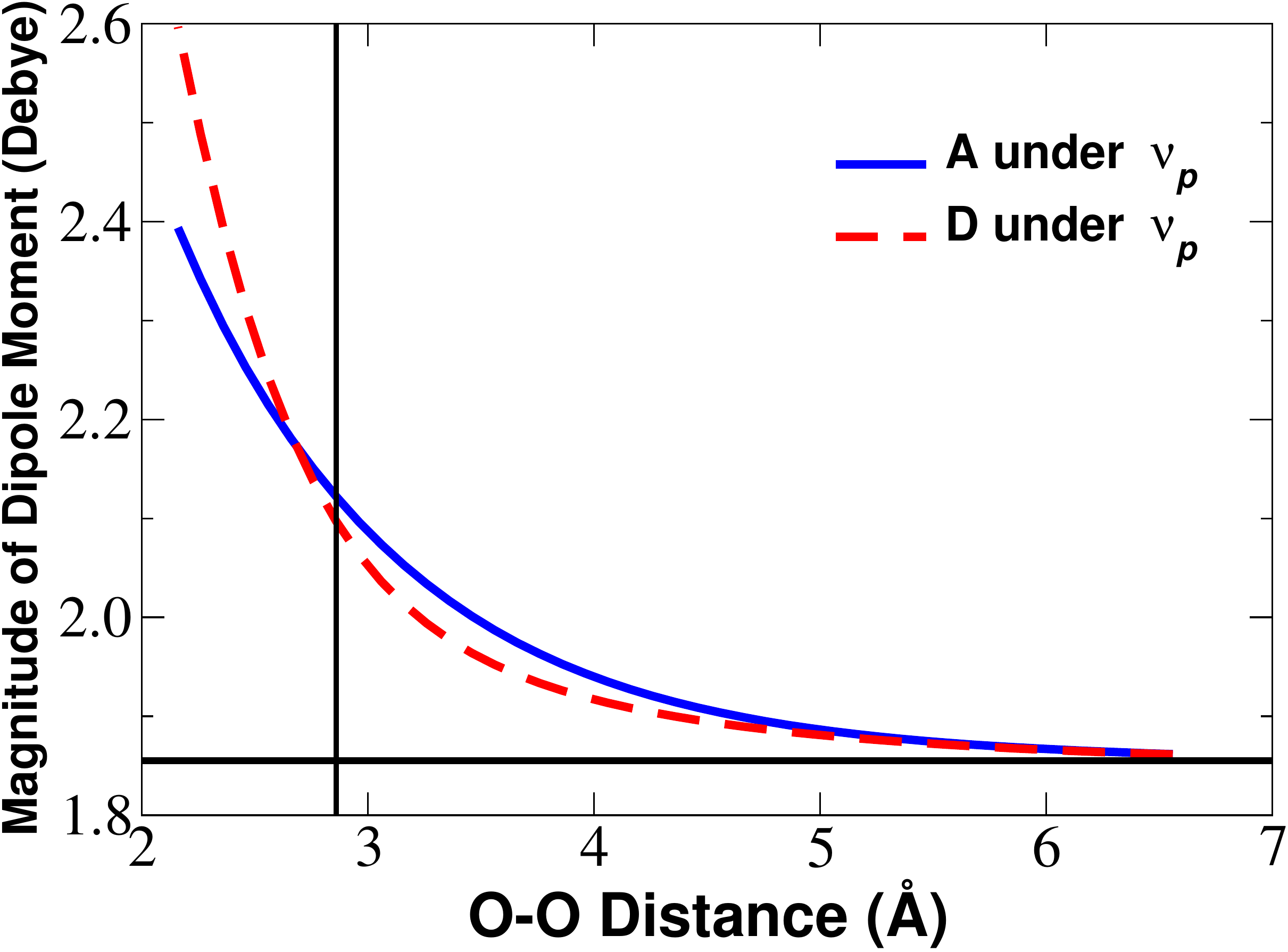} 
\caption{Dipole moments for the acceptor ($A$) and donor ($D$) water molecules in the $C_s$ water dimer as a function of the intermolecular distance. $v_p\left(\mathbf{r}\right)$ is expanded using Dunning's aug--cc--pVTZ basis set. The vertical line marks the equilibrium distance. The solid horizontal line shows the experimental dipole moment for the water monomer.}
\label{f:dm_along_scan}
\end{figure} 
\end{center}

\begin{figure}[H]\centering
\includegraphics[scale=0.30]{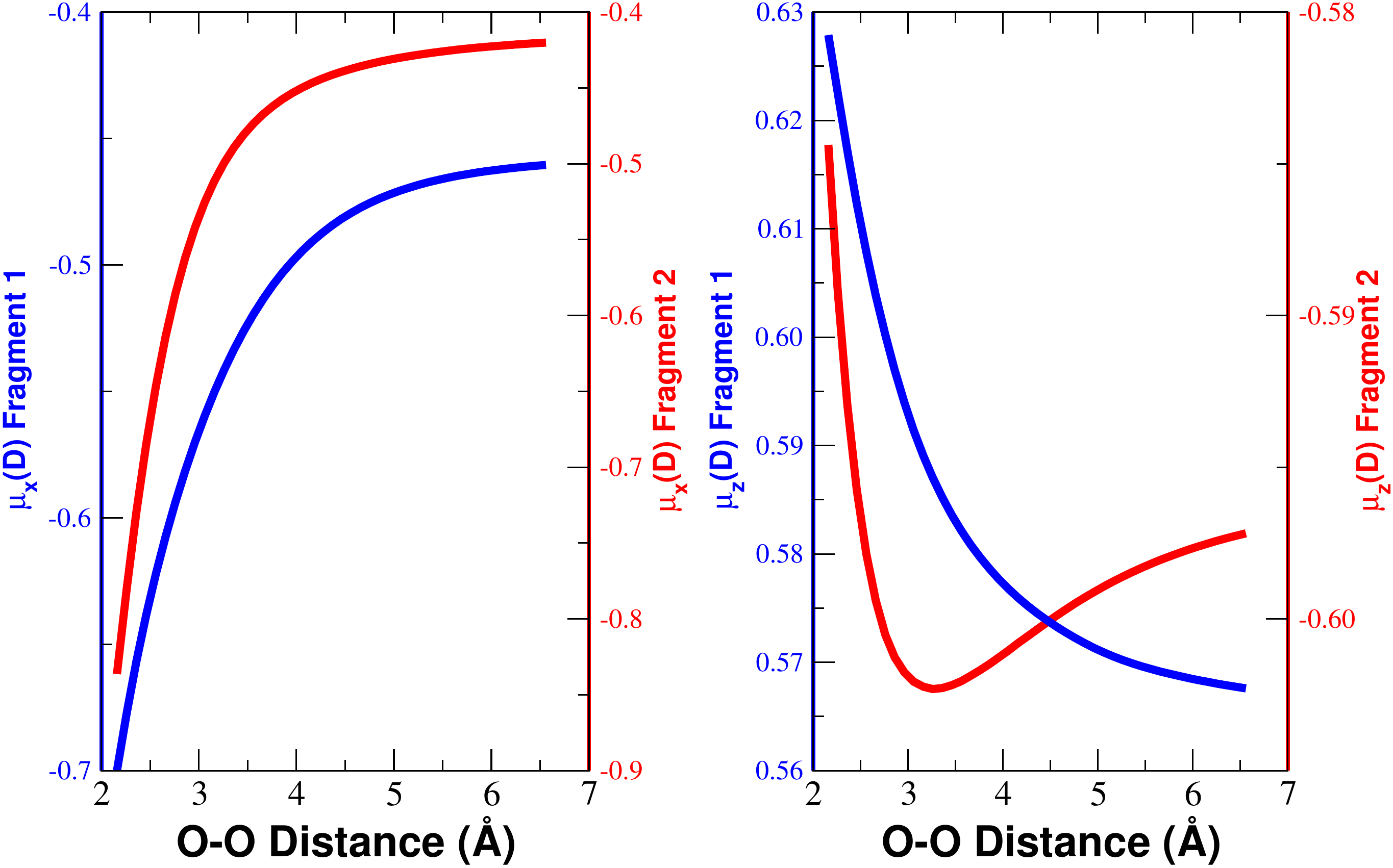}\\	
\includegraphics[scale=0.30]{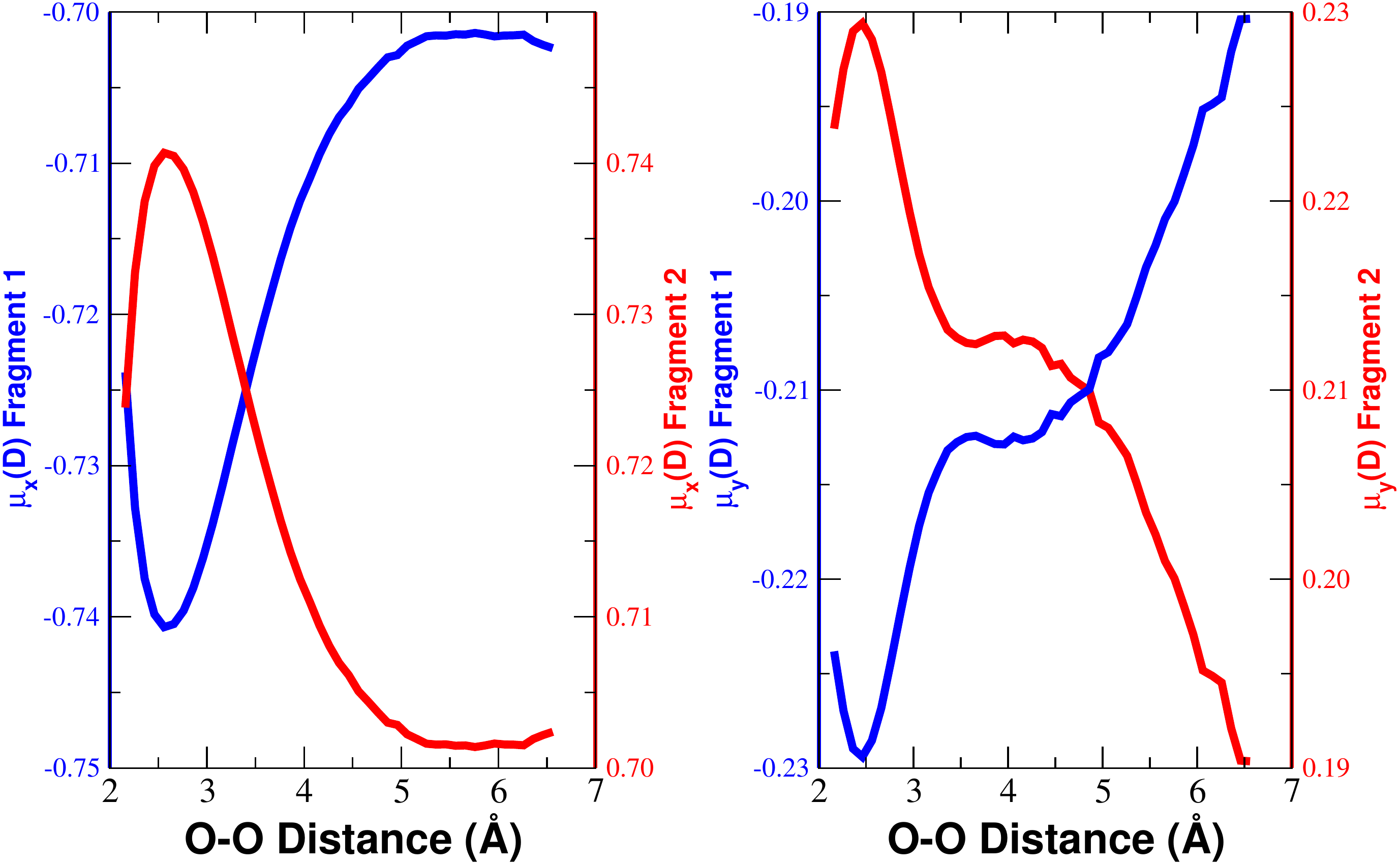}\\
\caption{Dipole moment components for the PDFT fragments in the $C_s$ (top panel) and in the $C_{2h}$ (bottom panel) water dimers at different intermolecular distances. The partition potential was expanded using Dunning's aug--cc--pVTZ basis set. Notice that the window of calculated dipole moments for the monomers in the $C_{2h}$ dimer is significantly smaller than in the $C_s$ structure.}
\label{f:dm_cs_c2h}
\end{figure}

We use our PDFT calculations to follow the evolution of the magnitudes of the fragment dipoles as a function of the separation between monomers. The results for the rigid scan of the $C_s$ dimer are plotted in Figure \ref{f:dm_along_scan}. At the equilibrium O--O distance ($\approx 2.86$ \AA) individual dipole moments are $\mu_A = 2.122$ and $\mu_D = 2.097$ Debyes. The changes in the dipole moments of the fragments at the equilibrium geometry are an indication that the distortions of the electron distributions needed for the hydrogen bond are properly described by the partition potential. It is also seen that in the range of intermolecular separation considered here, the dipole moment of the donor molecule increases to up to 2.597 Debyes in the repulsive region. As discussed above, the changes in the dipoles of the $C_{2h}$ dimer are significantly smaller than those of the $C_s$ dimer (Figure \ref{f:dm_cs_c2h}). This makes perfect sense because, for the $C_{2h}$ structure, the two hydrogen bonds point in opposite directions and thus the changes in monomer dipole moments cancel out. 

\begin{figure}\centering
\includegraphics[scale=0.60]{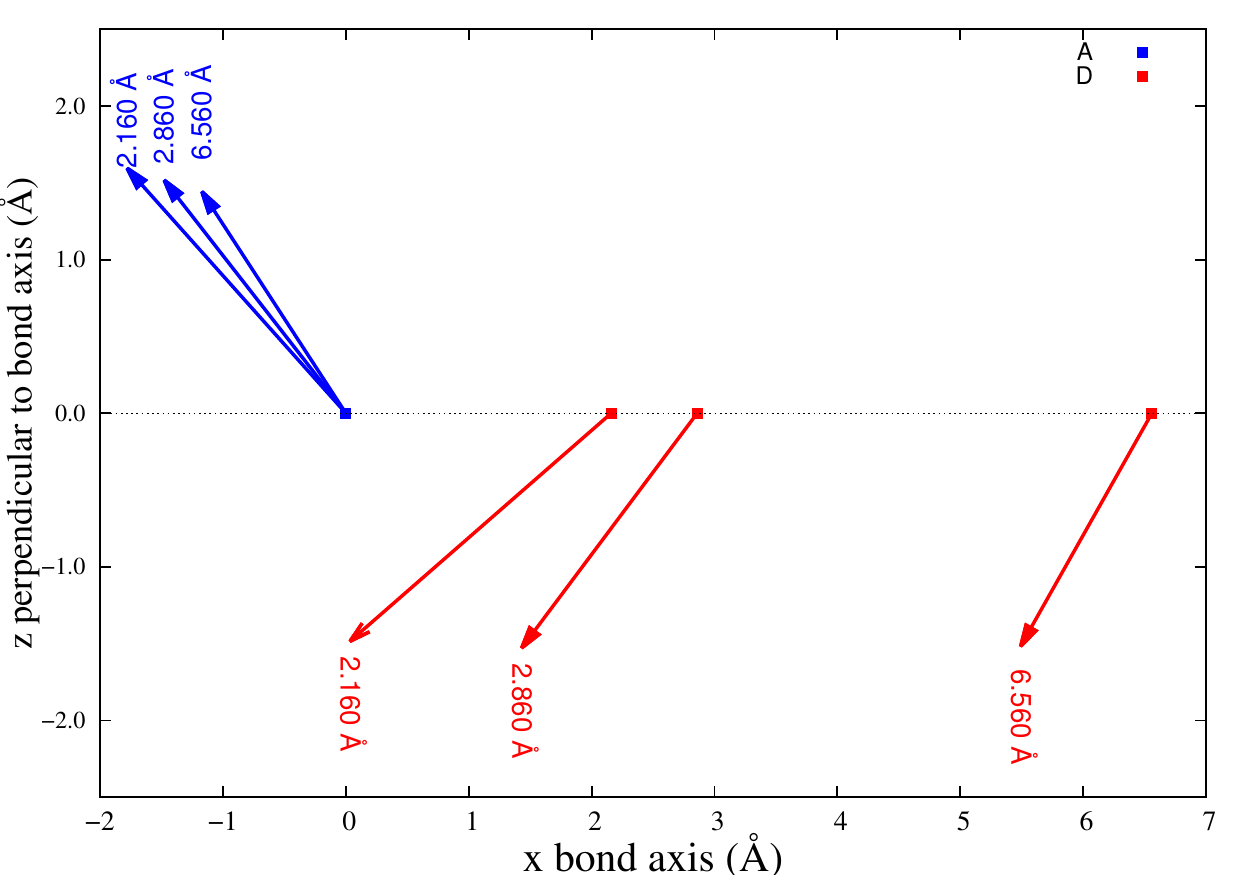}\\
\includegraphics[scale=0.60]{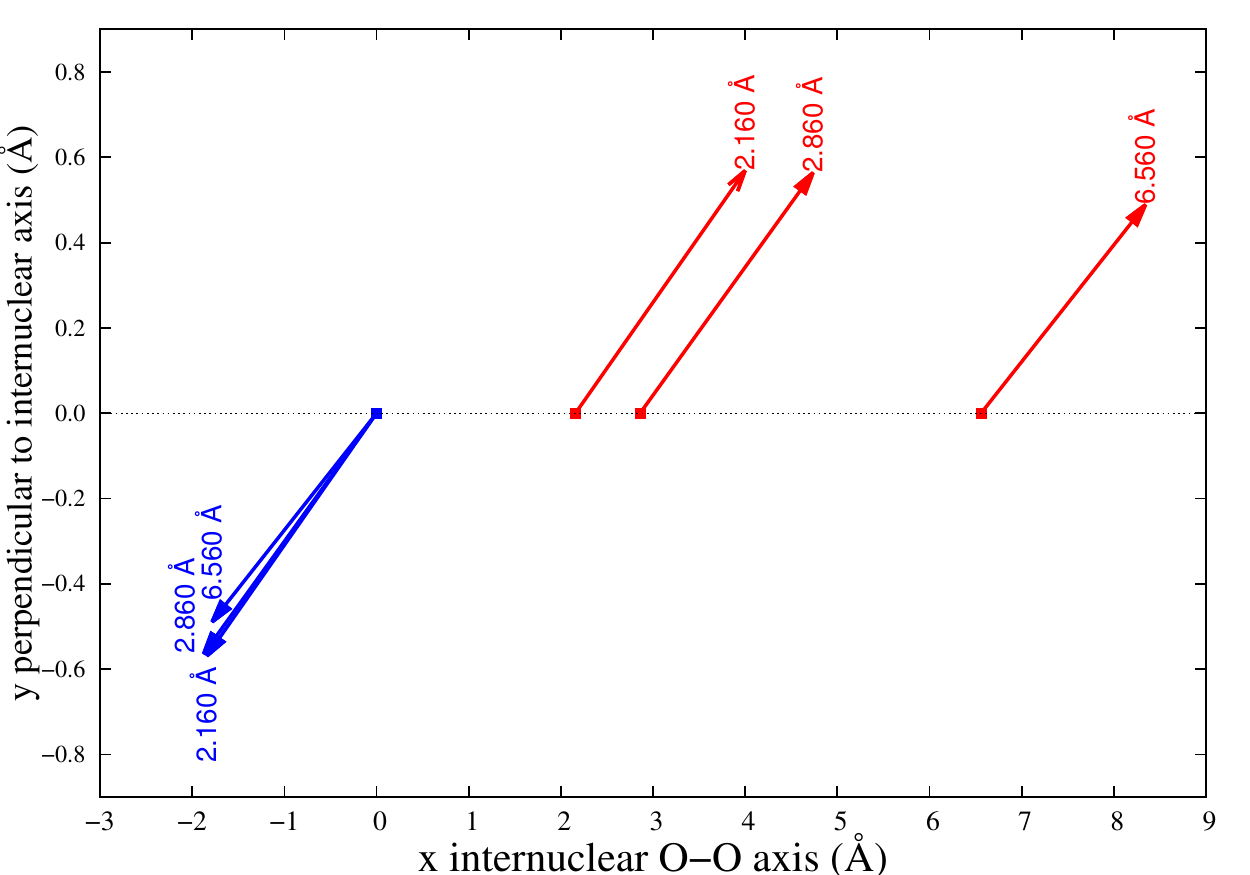}
\caption{Dipole moments for the acceptor ($A$, blue arrows, anti clockwise rotation) and donor ($D$, red arrows, clockwise rotation) fragments in the $C_s$ water dimer (top panel) and for both monomers in the $C_{2h}$ water dimer (bottom panel) at different intermolecular distances. The separation between fragments is provided at the tip of the arrows. For the $C_s$ case, the major changes happen at the $x$ component (the direction of the hydrogen bond) and $z$ component (the direction perpendicular to the hydrogen bond), and for the $C_{2h}$ water dimer the changes are seen in the $x$ and $y$ components.}
\label{f:dm_along_scan_vectors}
\end{figure}

Because of the intermolecular interaction, the dipole moments of the fragments change orientations as well as magnitudes. These changes are followed in Figure \ref{f:dm_along_scan_vectors} along the rigid scan of the $C_s$ dimer. As the fragments approach, both fragment have an increase in the x-componenent of the dipole moment following the classical O$^{\delta-}\cdots$H$^{\delta+}$--O$^{\delta-}$ electrostatic description of hydrogen bonds discussed in the introduction and depicted in the top panel of Figure \ref{f:cs_water_dimer}.

\section{\label{sec:concluding}Concluding Remarks}

PDFT shares many of the attractive features of density-based embedding methods \cite{JN14}. For example, it can be useful for QM/MM applications and force-field development. With approximations for the non-additive non-interacting kinetic energy functional, PDFT should also be amenable to efficient linear-scaling implementations. But what we have illustrated in this work is that PDFT can be employed to provide insightful chemical interpretations of the results beyond those that are possible with standard embedding: One can use PDFT to calculate the work involved in deforming isolated fragments to produce the unique fragments in the molecule.  We can calculate the dipoles of the fragments in the molecule, and interpret $v_p({\rm r})$ as a chemically significant reactivity potential \cite{CW07}, whose features can be meaningfully correlated with density distortions. Our results provide support for the orbital interaction picture of Reed and Weinhold \cite{weinhold_1988} for hydrogen-bond formation, but they do so without invoking orbitals.  

\section{\label{sec:acknowledgements}Acknowledgements}

We acknowledge support from the Universidad de Antioquia - Purdue grant No. PURDUE14-2-02.  J.N. and A.W. acknowledge support from the Office of Basic Energy Sciences, U.S. Department of Basic Energy Sciences, U.S. Department of Energy, under Grant No. DE-FG02-10ER16191. A.W. also acknowledges support from the Camille Dreyfus Teacher-Scholar Awards Program.

\section{\label{sec:refs}References}


\end{document}